\documentclass{intech}

\booktitle{Stochastic Control} 

\chaptertitle{Complexity and Stochastic Synchronization 
in Coupled Map Lattices and Cellular Automata} 

\authors{Ricardo L\'opez-Ruiz}
\affiliation{Universidad de Zaragoza}
\country{Spain}

\secondauthors{Juan R. S\'anchez} 
\secondaffiliation{Universidad Nacional de Mar del Plata}
\secondcountry{Argentine}

\def\sync{synchronization }

\begin{document}

\maketitle

\section{Introduction}

Nowadays the question {\it`what is complexity?'} is a challenge to be answered.
This question is triggering a great quantity of works in the frontier  
of physics, biology, mathematics and computer science.
Even more when this century has been told to be the century of {\it complexity} \citep{hawking}.
Although there seems to be no urgency to answer the above question,
many different proposals that have been developed to this respect 
can be found in the literature \citep{perakh}. In this context,
several articles concerning statistical complexity and stochastic processes are 
collected in this chapter.   

Complex patterns generated by the time evolution of a one-dimensional digitalized 
coupled map lattice are quantitatively analyzed in Section \ref{sec1}. 
A method for discerning complexity among the different 
patterns is implemented. The quantitative results indicate two zones in parameter space
where the dynamics shows the most complex patterns. These zones are located on the two edges 
of an absorbent region where the system displays spatio-temporal intermittency.

The \sync of two stochastically coupled one-dimensional 
cellular automata (CA) is analyzed in Section \ref{sec2}. 
It is shown that the transition to \sync is characterized by a dramatic increase of 
the statistical complexity of the patterns generated by the difference automaton.  
This singular behavior is verified to be present in several CA rules 
displaying complex behavior.

In Sections \ref{sec3} and \ref{sec4}, we are concerned with the stability analysis of patterns 
in extended systems. In general, it has been revealed to be a difficult task. 
The many nonlinearly interacting degrees of freedom can 
destabilize the system by adding small perturbations to some of them. The impossibility 
to control all those degrees of freedom finally drives the dynamics toward 
a complex spatio-temporal evolution. Hence, it is of a great interest to develop
techniques able to compel the dynamics toward a particular kind of structure. 
The application of such techniques forces the system to
approach the stable manifold of the required pattern, and
then the dynamics finally decays to that target pattern. 

Synchronization strategies in extended systems can be useful in order to achieve such goal.
In Section \ref{sec3}, we implement stochastic synchronization 
between the present configurations of a cellular automata and its precedent ones 
in order to search for constant patterns. In Section \ref{sec4}, 
this type of \sync is specifically used to find symmetrical patterns
in the evolution of a single automaton.

\section{Complexity in Two-Dimensional Patterns Generated by Coupled Map Lattices}
\label{sec1}

It should be kept in mind that in ancient epochs,
time, space, mass, velocity, charge, color, etc. were only perceptions.
In the process they are becoming concepts, different tools and instruments are invented for 
quantifying the perceptions. Finally, only with numbers the scientific laws emerge. 
In this sense, if by complexity it is to be understood that property present in all systems 
attached under the epigraph of `complex systems', this property should be reasonably quantified 
by the different measures that were proposed in the last years.
This kind of indicators is found in those fields where the concept of information
is crucial. Thus, the effective measure of complexity
\citep{grassberger} and the thermodynamical depth \citep{lloyd} come from physics 
and  other attempts  such as algorithmic complexity \citep{kolmogorov, chaitin}, 
Lempel-Ziv complexity \citep{lempel} and $\epsilon$-machine complexity
\citep{crutchfield} arise from the field of computational sciences.

In particular, taking into account the statistical properties of a system,
an indicator called the {\it LMC (L\'{o}pezRuiz-Mancini-Calbet) complexity}
has been introduced  \citep{lopezruiz94,lopezruiz95}. 
This magnitude identifies the entropy or information stored in a system and its disequilibrium
i.e., the distance from its actual state to the 
probability distribution of equilibrium, as the two basic ingredients for calculating
its complexity. If $H$ 
denotes the {\it information} stored in the system and $D$ is its 
{\it disequilibrium},
the LMC complexity $C$ is given by the formula:
\begin{eqnarray}
& C(\bar p)  =  H(\bar p)\cdot D(\bar p)\; = & \nonumber \\
   & =-k \left( \sum_{i=1}^N p_i\log p_i \right)\cdot
    \left( \sum_{i=1}^N \,\left( p_i - \frac{1}{N} \right)^2\right)\, &
    \label{eq:def-c}
\end{eqnarray}
where $\bar p=\{p_i\}$, with $p_i\geq 0$ and $i=1,\cdots,N$, represents the
distribution of the $N$ accessible states to the system, and $k$ is a constant
taken as $1/\log N$. 

As well as the Euclidean distance $D$ is present in the original LMC complexity,
other kinds of disequilibrium measures have been proposed in order to
remedy some statistical characteristics considered troublesome for
some authors \citep{feldman}.
In particular, some attention has been focused \citep{lin,martin} on 
the Jensen-Shannon divergence $D_{JS}$ as a measure for evaluating 
the distance between two different distributions $(\bar p_1,\bar p_2)$. 
This distance reads:
\begin{equation} 
D_{JS}(\bar p_1,\bar p_2) = H(\pi_1\bar p_1+\pi_2\bar p_2)-\pi_1H(\bar p_1)-\pi_2H(\bar p_2),
\label{eq:d-js}
\end{equation}
with $\pi_1,\pi_2$ the weights of the two probability distributions
$(\bar p_1,\bar p_2)$ verifying $\pi_1,\pi_2\geq 0$ and $\pi_1+\pi_2=1$.
The ensuing statistical complexity 
\begin{equation} 
C_{JS}=H\cdot D_{JS}
\label{eq:c-js}
\end{equation} 
becomes intensive and also keeps the property of distinguishing 
among distinct degrees of periodicity \citep{lamberti}. Here,
we consider $\bar p_2$ the equiprobability distribution and $\pi_1=\pi_2=0.5$.

\begin{figure}[t]
  \centering
  {\includegraphics[angle=0, width=10cm]{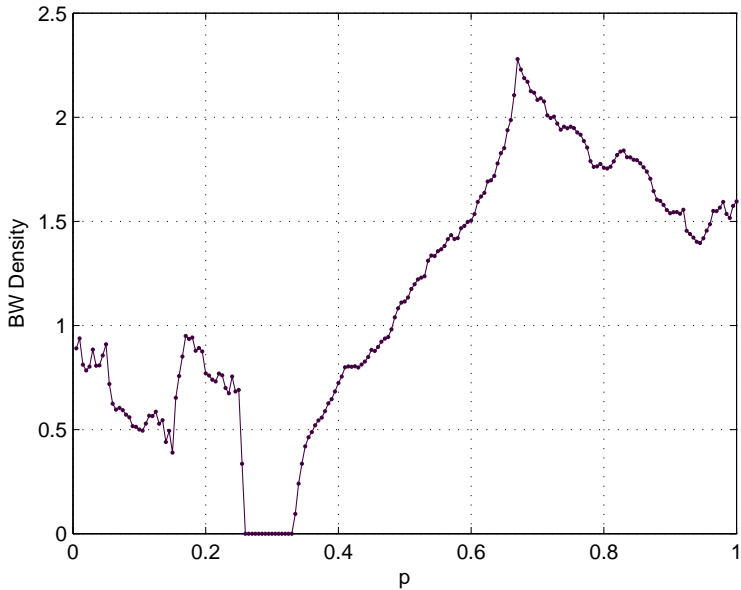}}
  \caption{$\beta$ versus $p$. The $\beta$-statistics (or BW density) 
  for each $p$ is the rate between the number of {\it black} and {\it white} cells 
  depicted by the system in the two-dimensional representation of its after-transient time evolution.
  (Computations have been performed with $\Delta p=0.005$ for a lattice of $10000$ sites after a transient 
  of $5000$ iterations and a running of other $2000$ iterations).}
  \label{fig1}
\end{figure}

As it can be straightforwardly seen, all these LMC-like complexities vanish 
both for completely ordered and for completely 
random systems as it is required for the correct asymptotic properties 
of a such well-behaved measure. Recently, they have been successfully used to
discern situations regarded as complex in discrete systems out of
equilibrium \citep{shiner,calbet,yuchen,rosso,rosso1,lovallo}. 

As an example, the local transition to chaos via
intermittency \citep{pomeau80} in the logistic map, $x_{n+1}=\lambda x_n(1-x_n)$
presents a sharp transition when C is plotted
versus the parameter $\lambda$ in the region around
the instability for $\lambda\sim \lambda_t=3.8284$. 
When $\lambda<\lambda_t$ the system approaches the laminar regime and 
the bursts become more unpredictable. The complexity increases. When the point
$\lambda=\lambda_t$ is reached a drop to zero occurs for the magnitude $C$.
The system is now periodic and it has lost its complexity.
The dynamical behavior of the system is finally well reflected in the magnitude $C$
(see \citep{lopezruiz95}).

When a one-dimensional array of such maps is put together a more complex behavior
can be obtained depending on the coupling among the units. Ergo the phenomenon
called {\it spatio-temporal intermittency} can emerge \citep{chate87,jensen90,jensen98}. 
This dynamical regime corresponds 
with a situation where each unit is weakly oscillating around a laminar state
that is aperiodically and strongly perturbed for a traveling burst. 
In this case, the plot of the one-dimensional lattice evolving in time 
gives rise to complex patterns on the plane. If the coupling among units
is modified the system can settle down in an absorbing phase where its dynamics 
is trivial \citep{coullet,toral} and then homogeneous patterns are obtained.
Therefore an abrupt transition to spatio-temporal intermittency can be depicted by
the system \citep{pomeau86,sinha} when modifying the coupling parameter.   

\begin{figure}[t]
  \centering
  {\includegraphics[angle=0, width=10cm]{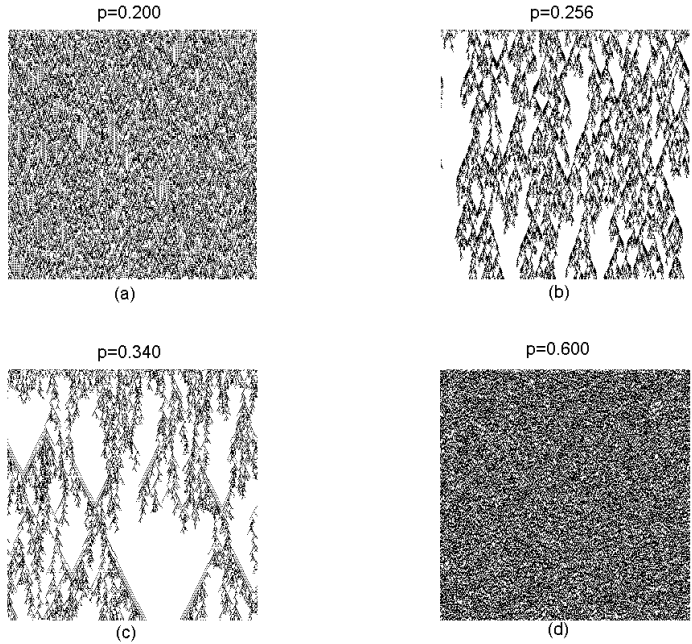}}
  \caption{Digitalized plot of the one-dimensional coupled map lattice (axe OX) evolving in time (axe OY)
  according to Eq. (\ref{eq:xn}): if $x_i^n>0.5$ the $(i,n)$-cell is put in white color 
  and if $x_i^n<0.5$ the $(i,n)$-cell is put in black color. The discrete time
  $n$ is reset to zero after the transitory. (Lattices of $300\times 300$ sites, 
  i.e., $0<i<300$ and $0<n<300$).}
  \label{fig2}
\end{figure}

In this section, we are concerned with measuring $C$ and $C_{JS}$ in a such
transition for a coupled map lattice of logistic type \citep{sanchez05a}.
Our system will be a line of sites, $i=1,\ldots,L$, with periodic
boundary conditions. In each site $i$ a local variable $x_i^{n}$ evolves 
in time ($n$) according to a discrete logistic equation. The interaction 
with the nearest neighbors takes place via a multiplicative coupling:
\begin{equation}
x_i^{n+1} = (4-3pX_i^{n})x_i^{n}(1-x_i^{n}),
\label{eq:xn}
\end{equation}  
where $p$ is the parameter of the system measuring the strength of the coupling ($0<p<1$).
The variable $X_i^{n}$ is the digitalized local mean field,
\begin{equation}
X_i^{n} = nint \left[\frac{1}{2}\: ({x_{i+1}^{n}+x_{i-1}^{n}}) \right] \: ,
\end{equation}
with {\it $nint(.)$} the integer function rounding its argument to the nearest integer.
Hence $X_i^{n}=0$ or $1$.

\begin{figure}[t]
  \centering
  {\includegraphics[angle=0, width=10cm]{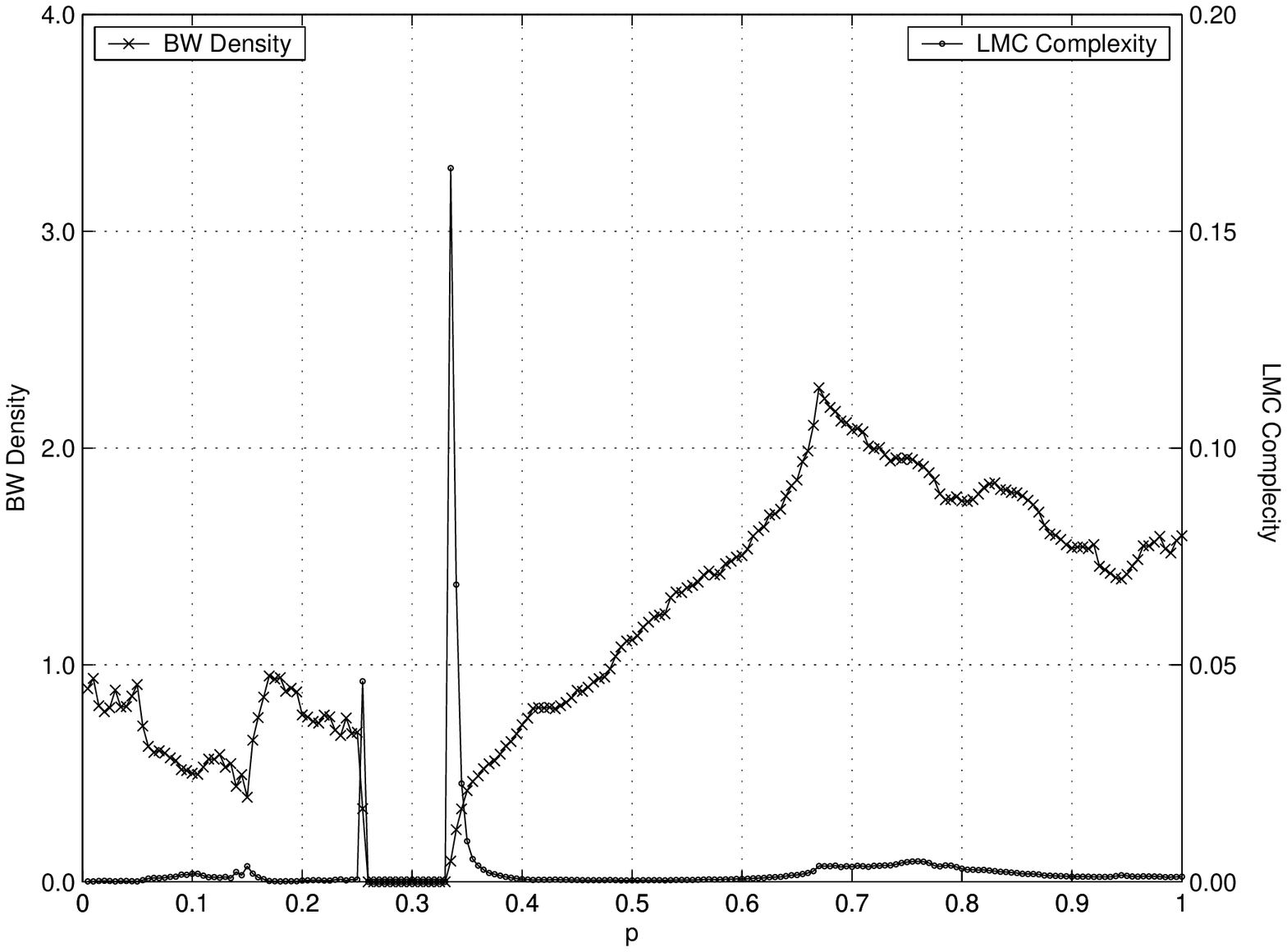}}
  \caption{($\bullet$) $C$ versus $p$. Observe the peaks of the LMC complexity located
  just on the borders of the absorbent region $0.258<p<0.335$,
  where $\beta=0$ ($\times$). 
  (Computations have been performed with $\Delta p=0.005$ for a lattice of $10000$ sites 
  after a transient of $5000$ iterations and a running of other $2000$ iterations).}
  \label{fig3}
\end{figure}

There is a biological motivation behind this kind of systems \citep{lopezruiz04,lopezruiz05}.
It could represent a {\it colony of interacting competitive individuals}.
They evolve randomly when they are independent ($p=0$). If some competitive interaction 
($p>0$) among them takes place the local dynamics loses its erratic component and becomes
chaotic or periodic in time depending on how populated the vicinity is.
Hence, for bigger $X_i^n$ more populated is the neighborhood of the individual $i$ and 
more constrained is its free action. At a first sight, it would seem that some particular 
values of $p$ could stabilize the system. In fact, this is the case. 
Let us choose a number of individuals for the colony ($L=500$ for instance),
let us initialize it randomly in the range $0<x_i<1$  and 
let it evolve until the asymptotic regime is attained.
Then the {\it black/white} statistics of the system is performed. That is,
the state of the variable $x_i$ is compared with the critical level $0.5$ for $i=1,\ldots,L$:
if $x_i>0.5$ the site $i$ is considered {\it white} (high density cell) and a counter $N_w$ is 
increased by one, or if $x_i<0.5$ the site $i$ is considered {\it black} (low density cell) and 
a counter $N_b$ is increased by one. This process is executed in the stationary regime
for a set of iterations. The {\it black/white} statistics is then the rate $\beta=N_b/N_w$.
If $\beta$ is plotted versus the coupling parameter $p$ the Figure \ref{fig1} is obtained.

The region $0.258<p<0.335$ where $\beta$ vanishes is remarkable. 
As stated above, $\beta$ represents the rate between the number of black cells and the number of white 
cells appearing in the two-dimensional digitalized representation of the colony evolution.
A whole white pattern is obtained for this range
of $p$. The phenomenon of spatio-temporal intermittency is displayed by the system 
in the two borders of this parameter region (Fig. \ref{fig2}). 
Bursts of low density (black color) travel in an
irregular way through the high density regions (white color). 
In this case two-dimensional complex patterns 
are shown by the time evolution of the system (Fig.  \ref{fig2}b-c). 
If the coupling $p$ is far enough from this region, i.e., $p<0.25$ or $p>0.4$,
the absorbent region loses its influence on the global dynamics and less structured 
and more random patterns than before are obtained (Fig.  \ref{fig2}a-d).   
For $p = 0$ we have no coupling of the maps, and each map generates so called fully developed
chaos, where the invariant measure is well-known to be symmetric around $0.5$. 
From this we conclude that $\beta(p = 0) = 1$. Let us observe that this symmetrical behavior
of the invariant measure is broken for small $p$, and $\beta$ decreases slightly 
in the vicinity of $p=0$.     

\begin{figure}[t]
  \centering
  {\includegraphics[angle=0, width=10cm]{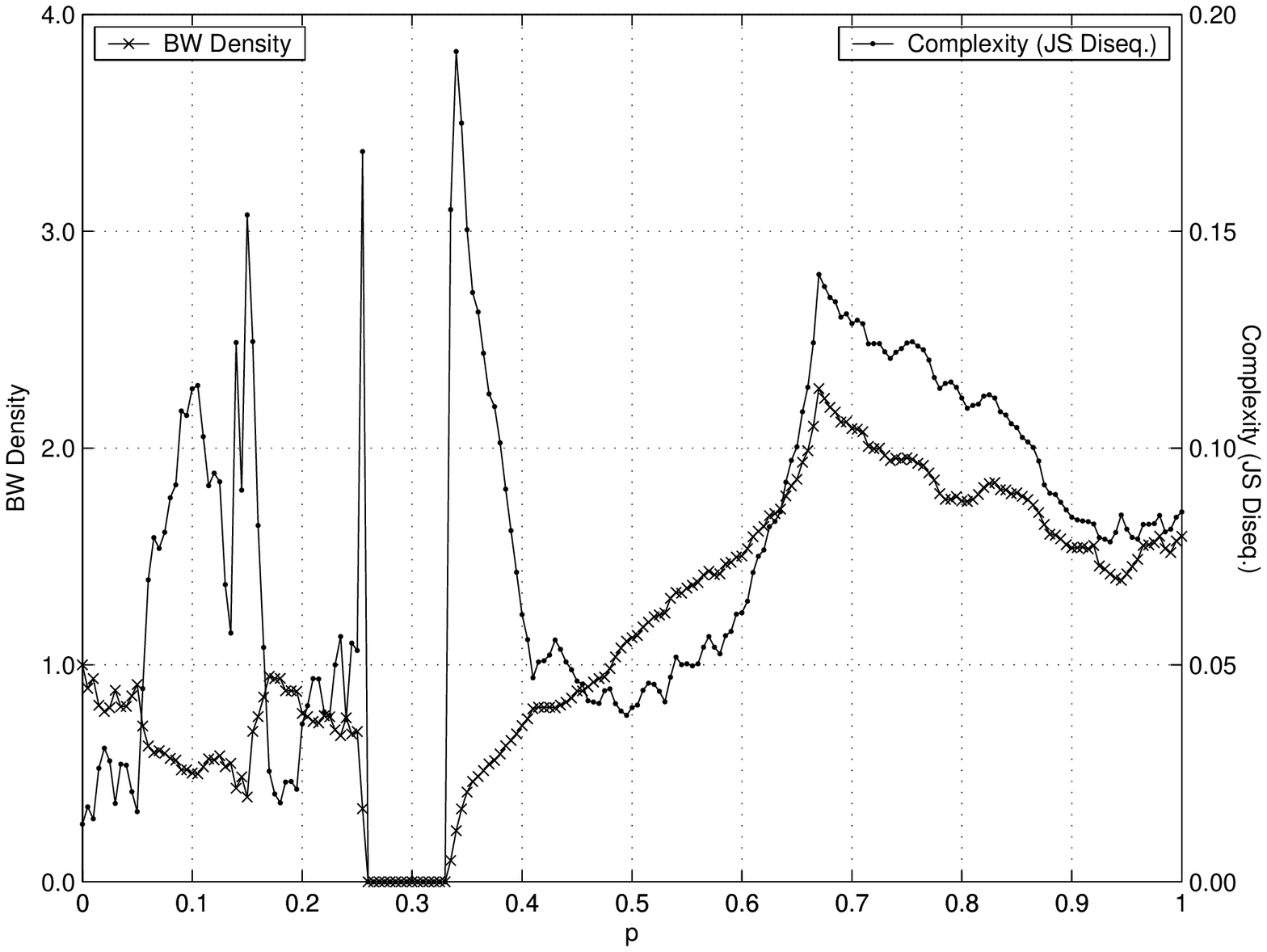}}
  \caption{($\cdot$) $C_{JS}$ versus $p$. The peaks of this modified LMC complexity
  are also evident just on the borders of the absorbent region $0.258<p<0.335$,
  where $\beta=0$ ($\times$). 
  (Computations have been performed with $\Delta p=0.005$ for a lattice of $10000$ sites 
  after a transient of $5000$ iterations and a running of other $2000$ iterations).}
  \label{fig4}
\end{figure}

If the LMC complexities are quantified as function of $p$, 
our {\it intuition} is confirmed.
The method proposed in \citep{lopezruiz95} to calculate $C$ is now adapted
to the case of two-dimensional patterns. First, 
we let the system evolve until the asymptotic regime is attained. 
This transient is discarded.
Then, for each time $n$, we map the whole lattice in a binary sequence:
$0$ if $x_i^n<0.5$ and $1$ if $x_i^n>0.5$, for $i=1,\ldots,L$.
This $L$-binary string is analyzed by blocks of $n_o$ bits, where
$n_o$ can be considered the scale of observation.
For this scale, there are $2^{n_o}$ possible states but only some of them are accessible.
These accessible states as well as their probabilities are found in the $L$-binary string.
Next, the magnitudes $H$, $D$, $D_{JS}$, $C$ and $C_{JS}$ are directly calculated for this 
particular time $n$ by applying the formulas (\ref{eq:def-c}-\ref{eq:c-js}). 
We repeat this process for a set of successive time units $(n,n+1,\cdots,n+m)$. 
The mean values of $H$, $D$, $D_{JS}$, $C$ and $C_{JS}$ for these $m$ time units
are finally obtained and plotted in Fig. \ref{fig3}-\ref{fig4}.  

Figures \ref{fig3},\ref{fig4} show the result for the case of $n_o=10$. 
Let us observe that the highest $C$ and $C_{JS}$ are reached when the dynamics displays 
spatio-temporal intermittency, that is, the {\it most complex patterns} 
are obtained for those values of $p$ that are located
on the borders of the absorbent region $0.258<p<0.335$.
Thus the plot of $C$ and $C_{JS}$ versus $p$ shows two tight peaks around the values
$p=0.256$ and $p=0.34$ (Fig. \ref{fig3},\ref{fig4}). Let us remark that  
the LMC complexity $C$ can be neglected far from the absorbent region.
Contrarily to this behavior, the magnitude $C_{JS}$ also shows high peaks
in some other sharp transition of $\beta$ located in the region $0<p<25$, and
an intriguing correlation with the {\it black/white} statistics in the region
$0.4<p<1$. All these facts as well as the stability study of the different dynamical 
regions of system (\ref{eq:xn}) are not the object of the present writing but
they deserve attention and a further study.  
 
If the detection of complexity in the two-dimensional case requires to identify 
some sharp change when comparing different patterns, 
those regions in the parameter space
where an abrupt transition happens should be explored 
in order to obtain the most complex patterns.
Smoothness seems not to be at the origin 
of complexity.  As well as a selected few distinct molecules 
among all the possible are in the basis of life \citep{mckay},
discreteness and its spiky appearance 
could indicate the way towards complexity.
Let us recall that the distributions 
with the highest LMC complexity are just those distributions  
with a spiky-like appearance \citep{calbet,plastino}. 
In this line, the striking result here exposed confirms the capability  of the 
LMC-like complexities for signaling a transition to complex behavior when regarding 
two-dimensional patterns \citep{sanchez05b}.

\section{Detecting Synchronization in Cellular Automata by Complexity Measurements}
\label{sec2}

Despite all the efforts devoted to understand the meaning of {\it complexity},
we still do not have an instrument in the laboratories specially designed for
quantifying this property. Maybe this is not the final objective of all those theoretical 
attempts carried out in the most diverse fields of knowledge in the last years
\citep{kolmogorov,chaitin,lempel,bennett,grassberger,lloyd,crutchfield,shiner},
but, for a moment, let us think in that possibility.

Similarly to any other device, our hypothetical apparatus will have an input
and an output. The input could be the time evolution of some variables 
of the system. The instrument records those signals, analyzes 
them with a proper program and finally screens the result in the form of 
a {\it complexity measurement}.
This process is repeated for several values of the 
parameters controlling the dynamics of the system.
If our interest is focused in the {\it most complex configuration} of the system 
we have now the possibility of tuning such an state by regarding the complexity 
plot obtained at the end of this process.

As a real applicability of this proposal, let us apply it to an {\it  \`a-la-mode} problem.
The clusterization or \sync of chaotic coupled elements was put in evidence at the beginning 
of the nineties \citep{kaneko,lopez91}. Since then, a lot of publications have been devoted
to this subject \citep{boccaletti}. Let us consider one particular of these systems to illuminate
our proposal.

(1) SYSTEM:
We take two coupled elementary one dimensional cellular
automata (CA: see next section in which CA are concisely explained)
displaying complex spatio-temporal dynamics \citep{wolfram}. 
It has been shown that this system can undergo through a \sync transition \citep{zanette}. 
The transition to full \sync occurs at a critical value $p_c$ of a \sync
parameter $p$. Briefly the numerical experiment
is as follows. Two $L$-cell CA with the same evolution rule $\Phi$ are started
from different random initial conditions for each automaton. Then, at each time
step, the dynamics of the coupled CA is governed by the successive
application of two evolution operators; the independent evolution of each CA according
to its corresponding rule $\Phi$ and the application of a stochastic operator that compares
the states $\sigma_i^1$ and $\sigma_i^2$ of all the cells, $i=1,...L$, in each
automaton. If $\sigma_i^1=\sigma_i^2$, both states are kept invariant. 
If $\sigma_i^1\neq\sigma_i^2$, they are left unchanged 
with probability $1-p$, but both states are updated either to $\sigma_i^1$
or to $\sigma_i^2$ with equal probability $p/2$.
It is shown in reference \citep{zanette} that there exists
a critical value of the \sync parameter ($p_c=0.193$ for the rule $18$) 
above for which full \sync is achieved.  

\begin{figure}[t]
\centering
  {\includegraphics[width=10cm]{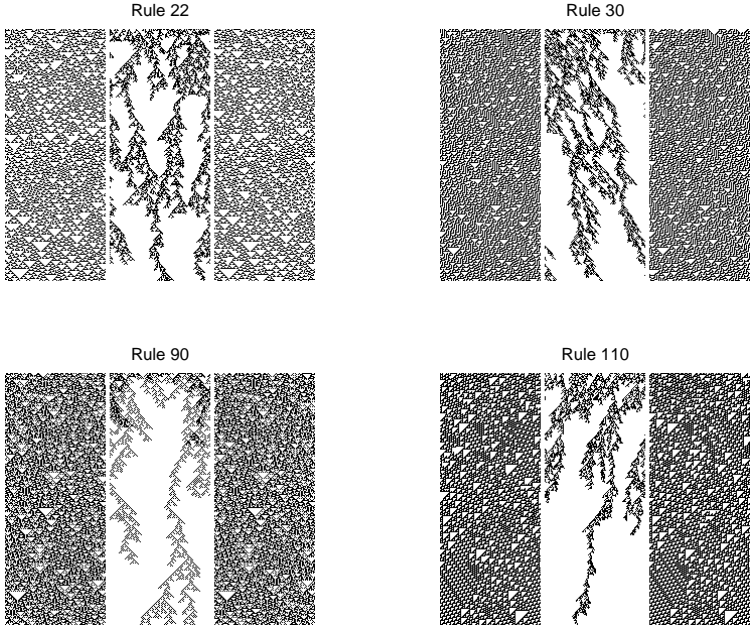}}
\caption{Spatio-temporal patterns just above the \sync transition.
The left and the right plots show $250$ successive states of the two coupled 
automata and the central plot is the corresponding difference automaton
for the rules $22$, $30$, $90$ and $110$. The number of sites is $L=100$
and the coupling probability is $p=0.23$.}
\label{fig5}
\end{figure}

(2) DEVICE: 
We choose a particular instrument to perform our measurements, 
that is capable of displaying the value of the {\it LMC complexity} ($C$) \citep{lopezruiz95}
defined as in Eq. (\ref{eq:def-c}), $C(\{\rho_i\}) =  H(\{\rho_i\})\cdot D(\{\rho_i\})\;$,
where $\{\rho_i\}$ represents the set of probabilities of the $N$ accessible 
discrete states of the system, 
with $\rho_i\geq 0$ , $i=1,\cdots,N$, and $k$ is a constant.
If $k=1/logN$ then we have the normalized complexity.
$C$ is a statistical measure of complexity that identifies the
entropy or information stored in a system and its disequilibrium,
i.e., the distance from its actual state to the 
probability distribution of equilibrium, as the two basic ingredients for calculating
the complexity of a system. 
This quantity vanishes both for completely ordered and for completely 
random systems giving then the correct asymptotic properties required for
a such well-behaved measure, and its calculation has been useful to successfully 
discern many situations regarded as complex. 

\begin{figure}[t]
\centering
{\includegraphics[width=10cm]{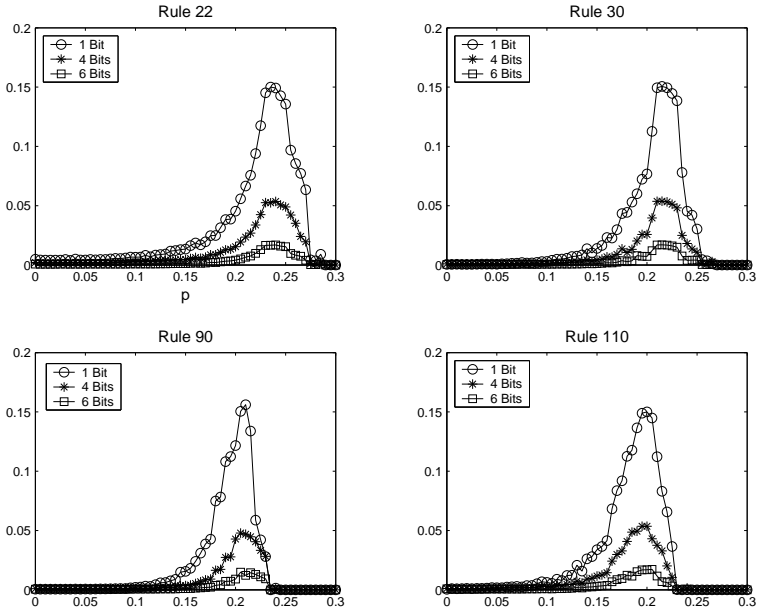}}
\caption{The normalized complexity $C$ versus the coupling probability $p$ 
  for different scales of observation: $n_o=1$ ($\circ$),$n_o=4$ ($\star$) 
  and $n_o=6$ ($\Box$). $C$ has been calculated over the last $300$ iterations 
  of a running of $600$ of them for a lattice with $L=1000$ 
  sites. The \sync transition is clearly depicted 
  around $p\approx 0.2$  for the different rules.}
\label{fig6}
\end{figure}

(3) INPUT:
In particular, the evolution of two coupled CA evolving under 
the rules $22$, $30$, $90$ and $110$ is analyzed. 
The pattern of the difference automaton will be the input of our device.
In Fig. \ref{fig5} it is shown for a coupling probability $p=0.23$,
just above the synchronization transition.
The left and the right plots show 
$250$ successive states of the two automata,
whereas the central plot displays the corresponding difference automaton.
Such automaton is constructed by comparing one by one all the sites ($L=100$)
of both automata and putting zero when the states $\sigma_i^1$
and $\sigma_i^2$, $i=1\ldots L$, are equal or putting one otherwise. 
It is worth to observe that 
the difference automaton shows an interesting {\it complex
structure} close to the \sync transition. This complex pattern
is only found in this region of parameter space. 
When the system if fully synchronized the difference automaton
is composed by zeros in all the sites, while when there is no \sync at all 
the structure of the difference automaton is completely random.

(4) METHOD OF MEASUREMENT:
How to perform the measurement of $C$ for such two-dimensional patterns has been 
presented in the former section \citep{sanchez05a}. 
We let the system evolve until the asymptotic regime is attained. 
The variable $\sigma_i^d$ in each cell of the difference pattern  
is successively translated to an unique binary sequence
when the variable $i$ covers the spatial dimension of the lattice, 
$i=1,\ldots,L$, and the time variable $n$ is consecutively increased.
This binary string is analyzed in blocks of $n_o$ bits, where
$n_o$ can be considered the scale of observation.
The accessible states to the system among the $2^{n_o}$ possible states 
is found as well as their probabilities.
Then, the magnitudes $H$, $D$ and $C$ are directly calculated
and screened by the device.   

(5) OUTPUT:
The results of the measurement are shown in Fig. \ref{fig6}.
The normalized complexity $C$ as a function of the \sync parameter $p$ is plotted for 
different coupled one-dimensional CA that evolve under the rules $22$, $30$, $90$ 
and $110$ , which are known to generate complex patterns. 
All the plots of Fig. \ref{fig6} were obtained using the following parameters: number of 
cell of the automata, $L=1000$; total evolution time, $T=600$ steps.
For all the cases and scales analyzed, the statistical complexity $C$
shows a dramatic increase close to the \sync transition. It reflects the complex structure
of the difference automaton and the capability of the measurement device here  
proposed for {\it clearly signaling}  the \sync transition of two coupled CA. 

These results are in agreement with the measurements of $C$
performed in the patterns generated by a one-dimensional logistic coupled map
lattice in the former section \citep{sanchez05a}. 
There the {\it LMC statistical complexity} ($C$)  
also shows a singular behavior close to the two edges 
of an absorbent region where the lattice displays spatio-temporal intermittency.
Hence, in our present case, the \sync region of the coupled systems can be interpreted
as an absorbent region of the difference system.
In fact, the highest complexity is reached on the border of this 
region for $p\approx 0.2$. The parallelism between both systems is therefore complete.

\section{Self-Synchronization of Cellular Automata}
\label{sec3}

Cellular automata (CA) are discrete dynamical systems, discrete both in space and time.  
The simplest one dimensional version of a cellular automaton
is formed by a lattice of $N$ sites or cells, 
numbered by an index $i=1,\ldots,N$, and with periodic boundary conditions. 
In each site, a local variable $\sigma_i$   
taking a binary value, either $0$ or $1$, is asigned.  The binary string 
$\sigma(t)$ formed by all sites values at time $t$ 
represents a configuration of the system.  The system evolves in time 
by the application of a rule $\Phi$. A new configuration $\sigma(t+1)$ is obtained
under the action of the rule $\Phi$ on the state $\sigma(t)$.
Then, the evolution of the automata can be writen as
\begin{equation}
\sigma(t+1) = \Phi\:[\sigma(t)].
\end{equation}
If coupling among nearest neighbors is used, 
the value of the site $i$, $\sigma_i(t+1)$, at time $t+1$ is a function 
of the value of the site itself at time $t$, $\sigma_i(t)$, and the values of 
its neighbors $\sigma_{i-1}(t)$ and $\sigma_{i+1}(t)$ at the same time.
Then, the local evolution is expressed as
\begin{equation}
\sigma_i(t+1) = \phi(\sigma_{i-1}(t),\sigma_i(t),\sigma_{i+1}(t)),
\end{equation}
being $\phi$ a particular realization of the rule $\Phi$.
For such particular implementation, there will be $2^3$ different local input 
configurations for each site and, for each one of them, a binary value can be 
assigned as output. Therefore there
will be $2^8$ different rules $\phi$, also called the {\it Wolfram rules}. 
Each one of these rules produces a 
different dynamical evolution.  In fact, dynamical behavior generated 
by all $256$ rules were already classified in four generic classes. 
The reader interested in the details of such classification is addressed to the 
original reference~\citep{wolfram}.  

CA provide us with simple dynamical systems, in which 
we would like to essay different methods 
of synchronization.  A stochastic synchronization technique was introduced 
in~\citep{zanette} that works in synchronizing 
two CA evolving under the same rule $\Phi$. The two CA are started 
from different initial conditions  
and they are supposed to have partial knowledge about each other.
In particular, the CA configurations, $\sigma^1(t)$ and $\sigma^2(t)$,  
are compared at each time step.
Then, a fraction $p$ of the total different sites are made equal (synchronized).  
The synchronization is stochastic since the location of the sites that are going to 
be equal is decided at random.
Hence, the dynamics of the two coupled CA, $\sigma(t)=(\sigma^1(t), \sigma^2(t))$, 
is driven by the successive application of two operators:  
\begin{enumerate}
	\item the deterministic operator given by the CA evolution rule $\Phi$,
	 $\Phi[\sigma(t)]=(\Phi[\sigma^1(t)], \Phi[\sigma^2(t)])$, and
	\item the stochastic operator $\Gamma_p$ that produces the result $\Gamma_p[\sigma(t)]$, 
	in such way that, if the sites are different ($\sigma_i^1\neq\sigma_i^2$), then $\Gamma_p$ 
	sets both sites equal to $\sigma_i^1$ with the probability $p/2$
	or equal to $\sigma_i^2$ with the same probability $p/2$.  
	In any other case $\Gamma_p$ leaves the sites unchanged.  
\end{enumerate}
Therefore the temporal evolution of the system can be written as
\begin{equation}
\sigma(t+1) = (\Gamma_p\circ\Phi)[\sigma(t)] = \Gamma_p[ (\Phi[\sigma^1(t)], \Phi[\sigma^2(t)])].
\end{equation}

\begin{figure}[t]
\centering
{\includegraphics[width=10cm]{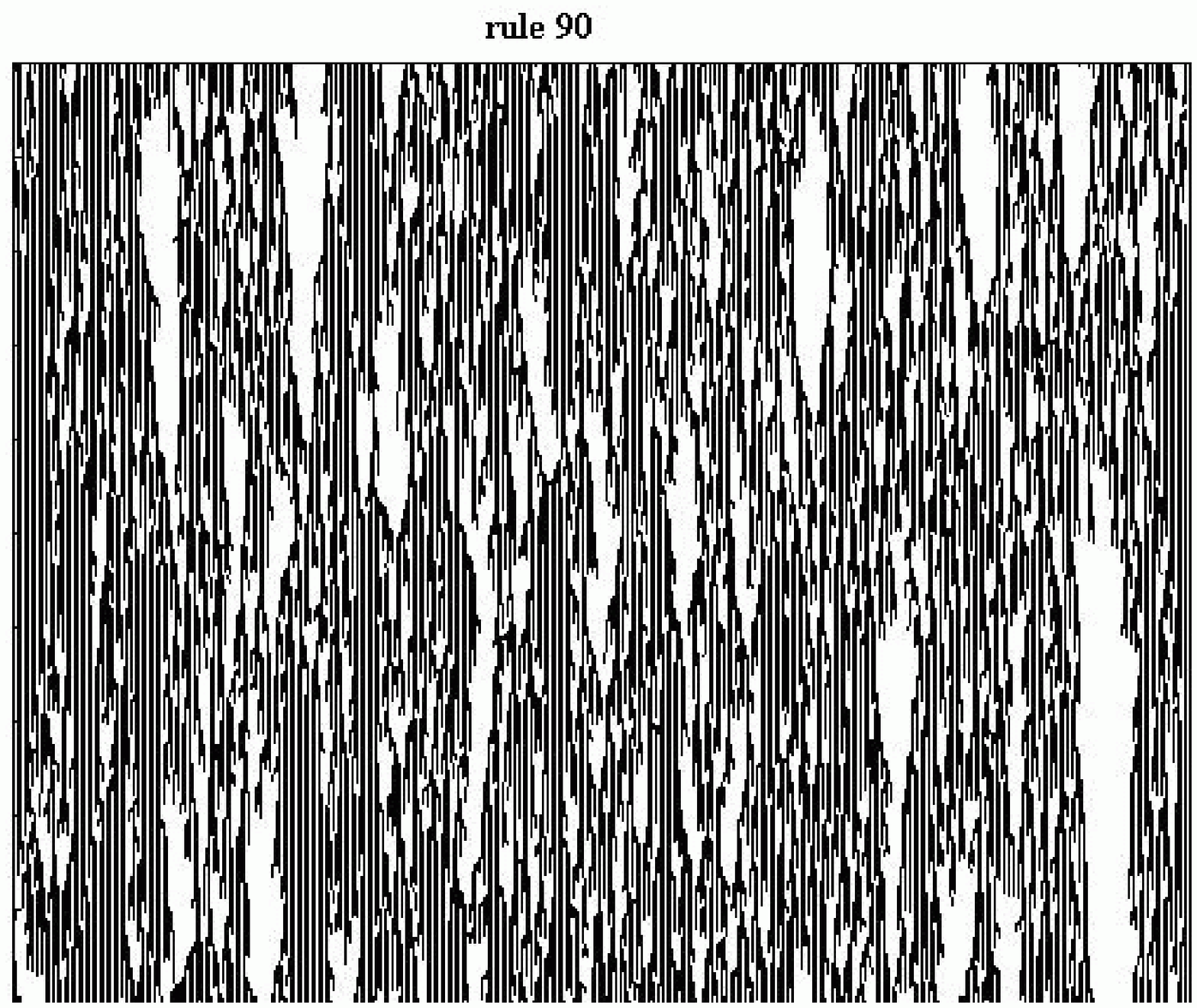}}
\caption{Rule $90$ has two stable patterns: one repeats the $011$ string
and the other one the $00$ string. Such patterns are reached by the first 
self-synchronization method but there is a dynamical competition between them.
In this case $p=0{.}9$.
Binary value $0$ is represented in white and $1$ in black.
Time goes from top to bottom.}
\label{fig7}
\end{figure}

A simple way to visualize the transition to synchrony can be done by displaying 
the evolution of the difference automaton (DA),
\begin{equation}
\delta_i(t)=\mid\sigma_i^1(t)-\sigma_i^2(t)\mid.
\end{equation}
The mean density of active sites for the DA
\begin{equation}
\rho(t)={1\over N}\sum_{i=1}^{N}\delta_i(t),
\end{equation}
represents the Hamming distance between the automata and 
verifies $0\leq\rho\leq 1$. 
The automata will be synchronized when $\lim_{t\rightarrow\infty}\rho(t)=0$. 
As it has been described in \citep{zanette} that
two different dynamical regimes, controlled by the parameter $p$, can be found 
in the system behavior: 
\begin{eqnarray*}
p<p_c & \rightarrow & \hbox{$\:\lim_{t\rightarrow\infty}\rho(t) \neq 0$ (no synchronization),} \\
p>p_c & \rightarrow & \hbox{$\:\lim_{t\rightarrow\infty}\rho(t)=0$ (synchronization)},
\end{eqnarray*}
being $p_c$ the parameter for which the transition to the synchrony occurs.  
When $p \lesssim p_c$ complex structures can be observed in the DA time evolution.  
In Fig. \ref{fig5}, typical cases of such behavior are shown near the synchronization transition. 
Lateral panels represent both CA evolving in time where the central strip displays the 
evolution of the corresponding DA.  When $p$ comes close to the critical value $p_c$ the evolution of
$\delta(t)$ becomes rare and resembles the problem of structures trying to percolate
in the plane~\citep{pomeau86}. A method to detect this kind of transition, based
in the calculation of a statistical measure of complexity for patterns, 
has been proposed in the former sections~\citep{sanchez05a},~\citep{sanchez05b}. 

\subsection{First Self-Synchronization Method}

Let us now take a single cellular automaton \citep{margolus,ilachinski}. 
If $\sigma^1(t)$ is the state of the automaton
at time t, $\sigma^1(t)=\sigma(t)$, and $\sigma^2(t)$ is the state obtained from the application 
of the rule $\Phi$ on that state, $\sigma^2(t)=\Phi[\sigma^1(t)]$, then the operator $\Gamma_p$ can be 
applied on the pair $(\sigma^1(t), \sigma^2(t))$, giving rise to the evolution law
\begin{equation}
	\sigma(t+1) = \Gamma_p[(\sigma^1(t), \sigma^2(t))] = \Gamma_p[ (\sigma(t), \Phi[\sigma(t)]) ].
	\label{eq-auto}
\end{equation}
The application of the $\Gamma_p$ operator is as follows.
When $\sigma_i^1\neq\sigma_i^2$, the sites $i$ of the state $\sigma^2(t)$ 
are updated to the correspondent values taken in $\sigma^1(t)$ with a probability $p$.
The updated array $\sigma^2(t)$ is the new state $\sigma(t+1)$. 

\begin{figure}[t]
\centering
{\includegraphics[width=10cm]{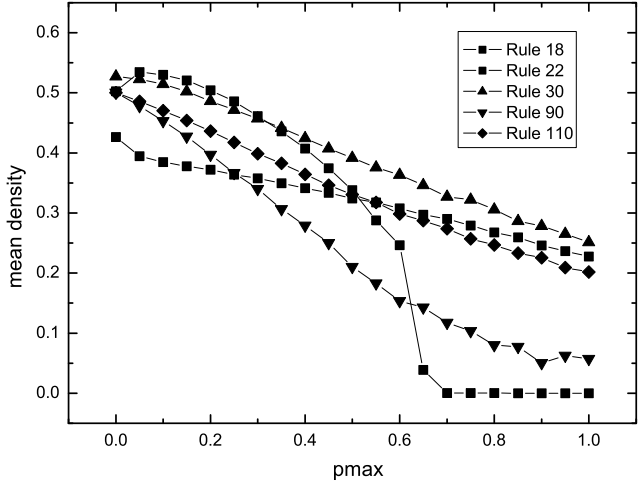}}
\caption{Mean density $\rho$ vs. $pmax=\tilde p$ for different rules evolving 
under the second synchronization method. The existence
of a transition to a synchronized state can be clearly observed for rule $18$.}
\label{fig8}
\end{figure}

It is worth to observe that if the system is initialized with
a configuration constant in time for the rule $\Phi$, $\Phi[\sigma]=\sigma$, 
then this state $\sigma$ is not modified when the dynamic equation (\ref{eq-auto})
is applied. Hence the evolution will produce
a pattern constant in time.      
However, in general, this stability is marginal. A small modification of the 
initial condition gives rise to patterns variable in time. 
In fact, as the parameter $p$ increases, 
a competition among the different marginally stable structures takes place.  
The dynamics drives the system to stay close to those states,
although oscillating continuously and randomly among them.
Hence, a complex spatio-temporal behavior is obtained.
Some of these patterns can be seen in Fig. \ref{fig7}.
However, in rule $18$, the pattern becomes stable and, independently
of the initial conditions, the system evolves toward this state, which
is the null pattern in this case \citep{sanchez06}.

\begin{figure}[t]
\centering
{\includegraphics[width=10cm]{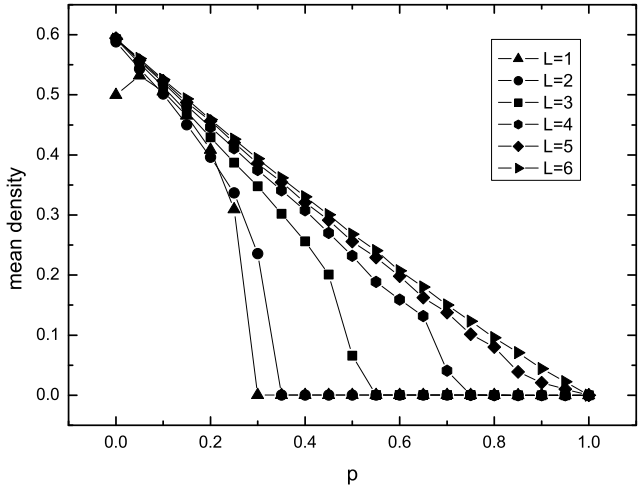}}
\caption{Mean density $\rho$ vs. $p$ for rule $18$ evolving under
the third self-synchronization method.
The existence of a transition to a synchronized state can be observed despite of
the randomness in the election of neighbors within a range $L$, up to $L=4$.}
\label{fig9}
\end{figure}

\subsection{Second Self-Synchronization Method}

Now we introduce a new stochastic element in the application of the operator $\Gamma_p$. 
To differentiate from the previous case we call it $\tilde\Gamma_{\tilde p}$.  
The action of this operator consists in applying at each time the operator 
$\Gamma_p$, with $p$ chosen at random in the interval $(0,\tilde p)$. 
The evolution law of the automaton is in this case: 
\begin{equation}
	\sigma(t+1) = \tilde\Gamma_{\tilde p}[(\sigma^1(t), \sigma^2(t))] = 
	\tilde\Gamma_{\tilde p}[ (\sigma(t), \Phi[\sigma(t)]) ].
	\label{eq-auto1}
\end{equation}

The DA density between the present state and the previous one, defined 
as $\delta(t)=\mid\sigma(t)-\sigma(t-1)\mid$, is plotted as a function of 
$\tilde p$ for different rules $\Phi$ in Fig. \ref{fig8}.
Only when the system becomes self-synchronized there will be a fall to zero in the DA density.
Let us observe again that the behavior reported in the first self-synchronization method 
is newly obtained in this case.  Rule $18$ undergoes a phase transition for a critical value of $\tilde p$. 
For $\tilde p$ greater than the critical value, the method is able to find the stable structure
of the system \citep{sanchez06}. For the rest of the rules the freezing phase is not found.
The dynamics generates patterns where the different marginally stable structures
randomly compete. Hence the DA density decays linearly with $\tilde p$
(see Fig. \ref{fig8}).

\begin{figure}[t]
\centering
{\includegraphics[width=10cm]{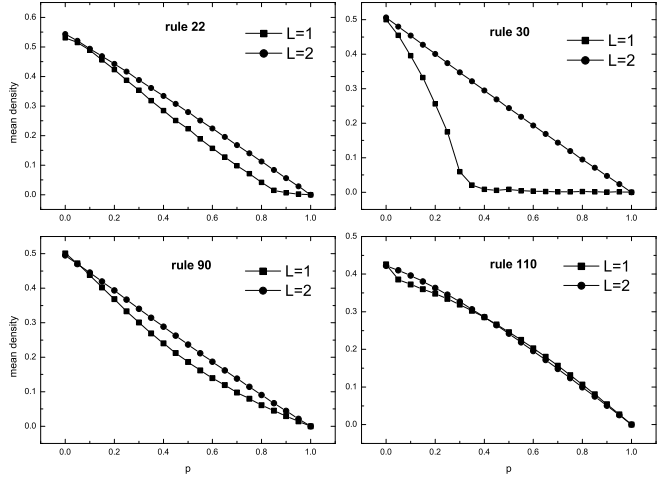}}
\caption{Mean density $\rho$ vs. $p$ for different rules evolving 
under the third self-synchronization method. 
The density of the system decreases linearly with $p$.}
\label{fig10}
\end{figure}

\subsection{Third Self-Synchronization Method}

At last, we introduce another type of stochastic element in the application of the rule $\Phi$. 
Given an integer number $L$, the surrounding of site $i$ at each time step is redefined. 
A site $i_l$ is randomly chosen among the $L$ neighbors of site $i$ to the left, $(i-L,\ldots,i-1)$.
Analogously, a site $i_r$ is randomly chosen among the $L$ neighbors of site $i$ to the right,
$(i+1,\ldots,i+L)$. The rule $\Phi$ is now applied on the site $i$ using the triplet $(i_l,i,i_r)$ 
instead of the usual nearest neighbors of the site. This new version of the rule is called $\Phi_L$,
being $\Phi_{L=1}=\Phi$.
Later the operator $\Gamma_p$ acts in identical way as in the first method.  
Therefore, the dynamical evolution law is:  
\begin{equation}
	\sigma(t+1) = \Gamma_p[(\sigma^1(t), \sigma^2(t))] = \Gamma_p[ (\sigma(t), \Phi_L[\sigma(t)]) ].
	\label{eq-auto2}
\end{equation}

The DA density as a function of $p$ is plotted in Fig. \ref{fig9} for the rule $18$
and in Fig. \ref{fig10} for other rules.
It can be observed again that the rule $18$ is a singular case that, 
even for different $L$, maintains the memory and continues to self-synchronize.
It means that the influence of the rule is even more important than the randomness 
in the election of the surrounding sites.
The system self-synchronizes and decays to the corresponding stable structure. 
Contrary, for the rest of the rules, 
the DA density decreases linearly with $p$ even for $L = 1$ as shown in Fig. \ref{fig10}. 
The systems oscillate randomly among their different marginally stable structures
as in the previous methods \citep{sanchez06}.

\section{Symmetry Pattern Transition in Cellular Automata with Complex Behavior}
\label{sec4}

\begin{figure}[t]
\centering
{\includegraphics[width=10cm]{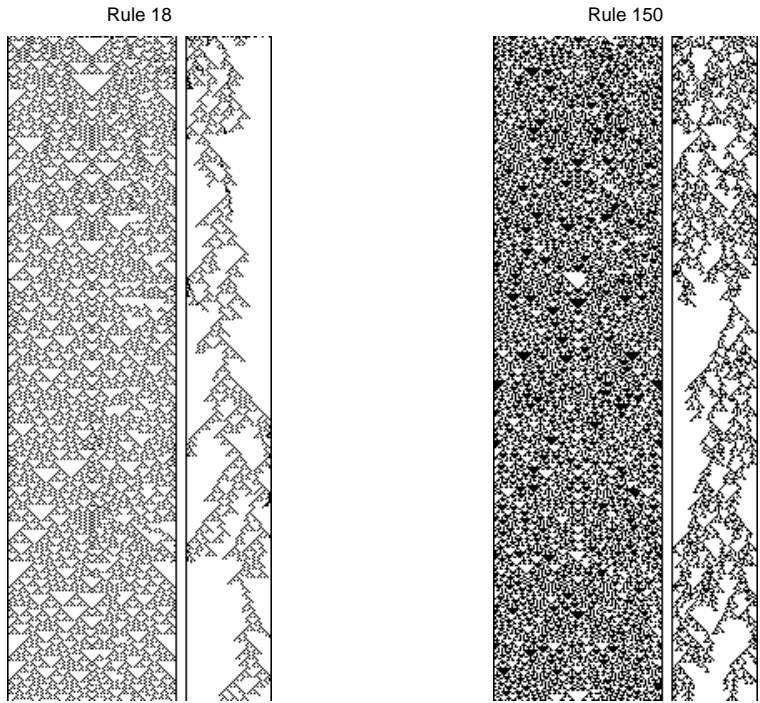}}
\caption{Space-time configurations of automata with $N=100$ sites iterated during 
$T=400$ time steps evolving under rules 18 and 150 for $p \lesssim p_c$. 
Left panels show the automaton evolution in time (increasing from top to bottom) 
and the right panels display the evolution of the corresponding DA.}
\label{fig11}
\end{figure}

\begin{figure}[]
\centering
{\includegraphics[height=12cm,width=10cm]{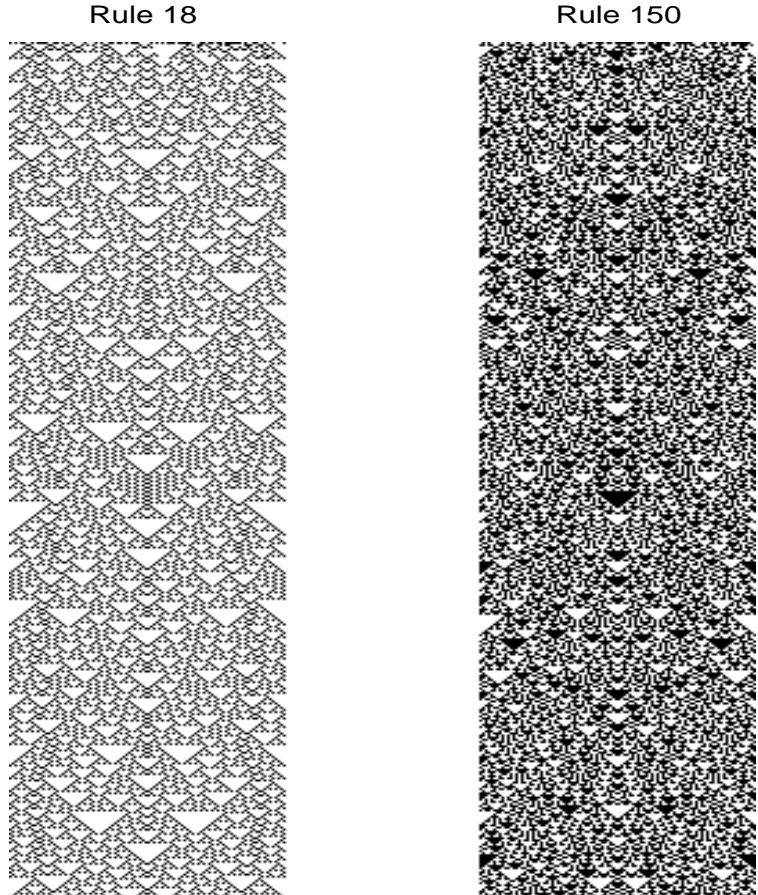}}
\caption{Time configurations of automata with $N=100$ sites iterated during $T=400$ 
time steps evolving under rules 18 and 150 using $p > p_c$. The space symmetry of 
the evolving patterns is clearly visible.}
\label{fig12}
\end{figure}

In this section, the stochastic synchronization method introduced in 
the former sections \citep{zanette} for two CA is specifically used to find symmetrical patterns
in the evolution of a single automaton. To achieve this goal the stochastic operator, below described,
is applied to sites symmetrically located from the center of the lattice. 
It is shown that a {\it symmetry} transition take place in the spatio-temporal pattern. 
The transition forces the automaton to evolve toward complex patterns that 
have mirror symmetry respect to the central axe of the pattern. In consequence,
this synchronization method can also be interpreted as a control technique for 
stabilizing complex symmetrical patterns.      

Cellular automata are extended systems, in our case one-dimensional strings 
composed of $N$ sites or cells. Each site is labeled by an index $i=1,\ldots,N$, 
with a local variable $s_i$ carrying a binary value, either $0$ or $1$.  
The set of sites values at time $t$ represents a configuration 
(state or pattern) $\sigma_t$ of the automaton.  During the automaton evolution, 
a new configuration $\sigma_{t+1}$ at time $t + 1$ is obtained by the 
application of a rule or operator $\Phi$ to the present configuration (see former section):
\begin{equation} \label{eq1}
\sigma_{t+1} = \Phi\:[\sigma_{t}]\:.
\end{equation}

\subsection{Self-Synchronization Method by Symmetry}

Our present interest \citep{sanchez08} resides in those CA evolving under rules capable
to show asymptotic complex behavior (rules of class III and IV). 
The technique applied here is similar to the synchronization scheme introduced by Morelli 
and Zanette~\citep{zanette} for two CA evolving under the same rule $\Phi$. 
The strategy supposes that the two systems have a {\it partial} knowledge one about 
each the other. At each time step and after the application of the rule $\Phi$,
both systems compare their present configurations $\Phi[\sigma^1_t]$ 
and $\Phi[\sigma^2_t]$ along all their extension 
and they synchronize a percentage $p$ of the total of their different sites.  
The location of  the percentage $p$ of sites that are going to be put equal is 
decided at random and, for this reason, it is said to be an stochastic synchronization.
If we call this stochastic operator $\Gamma_p$,
its action over the couple $(\Phi[\sigma^1_t],\Phi[\sigma^2_t])$ can be represented 
by the expression:
\begin{equation} \label{eq2}
(\sigma^1_{t+1},\sigma^2_{t+1}) = \Gamma_p(\Phi[\sigma^1_{t}],\Phi[\sigma^2_{t}])=
(\Gamma_p\circ\Phi)(\sigma^1_{t},\sigma^2_{t}).
\end{equation}  

\begin{figure}[t]
\centering
{\includegraphics[width=10cm]{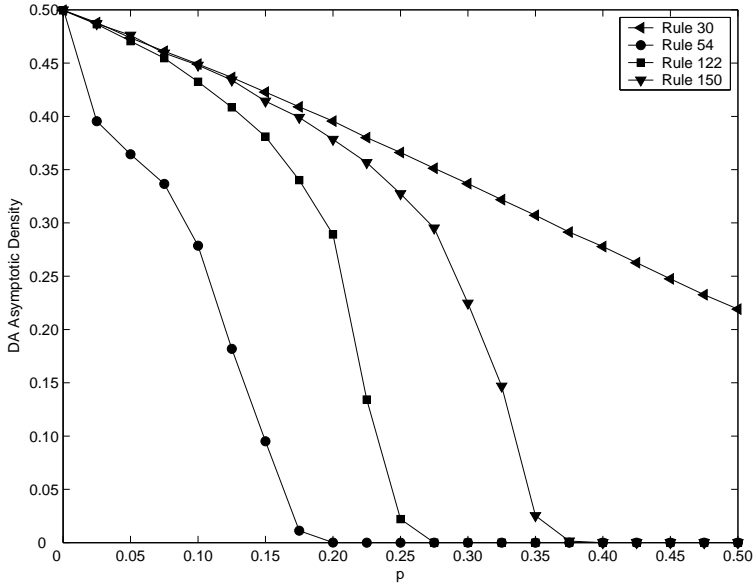}}
\caption{Asymptotic density of the DA for different rules is plotted as a 
function of the coupling probability $p$. Different values of $p_c$ for each 
rule appear clearly at the points where $\rho \to 0$. The automata with $N=4000$ 
sites were iterated during $T=500$ time steps. The mean values of the last 
$100$ steps were used for density calculations.}
\label{fig13}
\end{figure}

The same strategy can be applied to a single automaton with a even 
number of sites \citep{sanchez08}. 
Now the evolution equation, $\sigma_{t+1} = (\Gamma_p\circ\Phi)[\sigma_t]$, 
given by the successive action of the two operators $\Phi$ and $\Gamma_p$, 
can be applied to the configuration $\sigma_t$ as follows:  
\begin{enumerate}
	\item the deterministic operator $\Phi$ for the evolution of the automaton
	produces $\Phi[\sigma_t]$, and,
	\item the stochastic operator $\Gamma_p$, produces the result 
	$\Gamma_p(\Phi[\sigma_t])$, in such way that, if sites symmetrically located 
	from the center are different, i.e. $s_{i} \neq s_{N-i+1}$, then 
	$\Gamma_p$ equals $s_{N-i+1}$ to $s_i$ with probability $p$. 
	$\Gamma_p$ leaves the sites unchanged with probability $1-p$.  
\end{enumerate}

\begin{table}[]
 		\begin{tabular}{|c|c|c|c|c|c|c|c|c|c|c|c|c|c|}
		    \hline
 			Rule  & $18$ & $22$ & $30$ & $54$ & $60$ & $90$ & $105$ & $110$ & 
			$122$ & $126$ & $146$ & $150$ & $182$ \\
 			\hline 		
 			$p_c$ & $0.25$ & $0.27$ & $1.00$ & $0.20$ & $1.00$ & $0.25$ & 
			$0.37$ & $1.00$ & $0.27$ & $0.30$ & $0.25$ & $0.37$ & $0.25$ \\
			\hline
 		\end{tabular}
\caption{Numerically obtained values of the critical probability $p_c$ 
for different rules displaying complex behavior. Rules that can not sustain
symmetric patterns need fully coupling of the symmetric sites, i.e. ($p_c=1$).}
\label{table1}
\end{table}

A simple way to visualize the transition to a symmetric pattern can be 
done by splitting the automaton in two subsystems ($\sigma^1_t,\sigma^2_t$), 
\begin{itemize}
	\item $\sigma^1_t$, composed by the set of sites $s(i)$ with $i=1,\ldots,N/2$ and
	\item $\sigma^2_t$, composed the set of symmetrically located sites $s(N-i+1)$ with $i=1,\ldots,N/2$, 
\end{itemize}
and displaying the evolution of the difference automaton (DA), defined as
\begin{equation}
\delta^{t}=\mid\sigma_t^1 - \sigma_t^2 \mid \:.
\end{equation}
The mean density of active sites for the difference automaton, defined as
\begin{equation}
\rho^t={2\over N}\sum_{i=1}^{N/2}\delta^t
\end{equation}
represents the Hamming distance between the sets $\sigma^1$ and $\sigma^2$. 
It is clear that the automaton will display a symmetric pattern when 
$\lim_{t\rightarrow\infty}\rho^t=0$. 
For class III and IV rules, a symmetry transition controlled by the parameter $p$ is found.
The transition is characterized by the DA behavior:
\begin{eqnarray*}
\hbox{when $p<p_c$} & \rightarrow & \hbox{$\:\lim_{t\to\infty}\rho^t \neq 0$ (complex non-symmetric patterns),} \\
\hbox{when $p>p_c$} & \rightarrow & \hbox{$\:\lim_{t\to\infty}\rho^t=0$ (complex symmetric patterns)}.
\end{eqnarray*}
The critical value of the parameter $p_c$ signals the transition point.  
 
In Fig. \ref{fig11} the space-time configurations of automata 
evolving under rules 18 and 150 are shown for $p \lesssim p_c$. 
The automata are composed by $N=100$ sites and were iterated during $T=400$ time steps.
Left panels show the automaton evolution in time (increasing from top to bottom) and the 
right panels display the evolution of the corresponding DA. For $p \lesssim p_c$, 
complex structures can be observed in the evolution of the DA. As $p$ approaches 
its critical value $p_c$, the evolution of the DA become more stumped and reminds 
the problem of structures trying to percolate the plane~\citep{pomeau86,sanchez05a}.
In Fig. \ref{fig12} the space-time configurations of the same automata are displayed for $p > p_c$. 
Now, the space symmetry of the evolving patterns is clearly visible.

Table \ref{table1} shows the numerically obtained values of $p_c$ 
for different rules displaying complex behavior. It can be seen that some rules can not sustain
symmetric patterns unless those patterns are forced to it by fully coupling 
the totality of the symmetric sites ($p_c=1$).  
The rules whose local dynamics verify $\phi(s_1,s_0,s_2)=\phi(s_2,s_0,s_1)$
can evidently sustain symmetric patterns, and these structures are induced 
for $p_c<1$ by the method here explained. 

Finally, in Fig. \ref{fig13} the asymptotic density of the DA, $\rho^t$ for $t \to \infty$, 
for different rules is plotted as a function of the coupling probability $p$. 
The values of $p_c$ for the different rules appear clearly at the points where $\rho \to 0$.

\section{Conclusion}

A method to measure statistical complexity in extended systems has been implemented.
It has been applied to a transition to spatio-temporal complexity 
in a coupled map lattice and to a transition to \sync in two stochastically
coupled cellular automata (CA). The statistical indicator shows a peak 
just in the transition region, marking clearly the change of dynamical behavior 
in the extended system. 

Inspired in stochastic synchronization methods for CA, different schemes 
for self-synchronization of a single automaton have also been proposed and analyzed.  
Self-synchronization of a single automaton can be interpreted 
as a strategy for searching and controlling the structures of the system
that are constant in time. In general, it has been found that a competition 
among all such structures is established,
and the system ends up oscillating randomly among them. 
However, rule $18$ is a unique position among all rules because,
even with random election of the neighbors sites, 
the automaton is able to reach the configuration constant in time.

Also a transition from asymmetric to symmetric patterns in time-dependent 
extended systems has been described.
It has been shown that one dimensional cellular automata, 
started from fully random initial conditions, 
can be forced to evolve into complex {\it symmetrical} patterns by stochastically 
coupling a proportion $p$ of pairs of sites located 
at equal distance from the center of the lattice. 
A nontrivial critical value of $p$ must be surpassed in order to obtain symmetrical patterns 
during the evolution. This strategy could be used as an alternative to classify the cellular 
automata rules -with complex behavior- between those that support time-dependent symmetric patterns 
and those which do not support such kind of patterns.


\end{document}